\documentclass[a4paper,11pt]{article}
\pdfoutput=1 

\usepackage{jheppub} 

\usepackage[T1]{fontenc} 
\usepackage{arydshln}
\usepackage{slashed}
\usepackage{multirow}
\usepackage{bm}
\usepackage{amsmath}
\usepackage{amssymb}

\def\a{\alpha}

\def\e{\epsilon}

\def\s{\sigma}

\def\ph{\phi}

\def\th{\theta}

\def\mO{\mathcal{O}}

\makeatletter
\newsavebox\myboxA
\newsavebox\myboxB
\newlength\mylenA
\newcommand*\xoverline[2][0.75]{%
    \sbox{\myboxA}{$\m@th#2$}%
    \setbox\myboxB\null
    \ht\myboxB=\ht\myboxA%
    \dp\myboxB=\dp\myboxA%
    \wd\myboxB=#1\wd\myboxA
    \sbox\myboxB{$\m@th\overline{\copy\myboxB}$}
    \setlength\mylenA{\the\wd\myboxA}
    \addtolength\mylenA{-\the\wd\myboxB}%
    \ifdim\wd\myboxB<\wd\myboxA%
       \rlap{\hskip 0.5\mylenA\usebox\myboxB}{\usebox\myboxA}%
    \else
        \hskip -0.5\mylenA\rlap{\usebox\myboxA}{\hskip 0.5\mylenA\usebox\myboxB}%
    \fi}
\makeatother

\newcommand{\be}{\begin{eqnarray}}
\newcommand{\ee}{\end{eqnarray}}
\def\a{\alpha}

\title{
Independently Parameterised Momenta Variables and Monte Carlo IR Subtraction
}

\author[a]{Peter Cox}

\author[a]{and Tom Melia}

\affiliation[a]{Kavli Institute for the Physics and Mathematics of the Universe (WPI), The University of Tokyo Institutes for Advanced Study, University of Tokyo, Kashiwa 277-8583, Japan}

\emailAdd{peter.cox@ipmu.jp}
\emailAdd{tom.melia@ipmu.jp}

\preprint{IPMU18-0141}

\abstract{
We introduce a system of parameters for the Monte Carlo generation of  Lorentz invariant phase space that is particularly well-suited to the treatment of the infrared divergences that occur in the most singular, Born-like configurations of $1\to n$ QCD processes. A key feature is that particle momenta are generated independently of one another, leading to a simple parameterisation of all such IR limits. We exemplify the use of these variables in conjunction with the projection to Born subtraction technique  at next-to-next-to-leading order. The geometric origins of this parameterisation lie in a coordinate chart on a Grassmannian manifold.
}

\notoc

\begin{document} 
\maketitle
\flushbottom
\newpage

\section{Introduction}

Differential predictions for generic collider observables rely on a Monte Carlo (MC) sampling of Lorentz invariant phase space,
\be
d\Phi_n =\frac{d^{d-1}{\bf p}_1}{(2\pi)^{d-1}2E_1} \cdots  \frac{d^{d-1}{\bf p}_n}{(2\pi)^{d-1}2E_n}  \,(2\pi)^d\delta^{(d)}(P-\sum_i p_i)  \,   ,
\label{eq:phspcanonical}
\ee
with a prescription for dealing with the  infrared (IR) singularities that arise from soft and collinear limits of amplitudes. 
A number of state-of-the-art techniques exist to obtain next-to-next-to-leading order (NNLO) precision in QCD for exclusive observables~\cite{GehrmannDeRidder:2005cm,GehrmannDeRidder:2005aw,GehrmannDeRidder:2005hi,Daleo:2006xa,Daleo:2009yj,Gehrmann:2011wi,Boughezal:2010mc,GehrmannDeRidder:2012ja,Currie:2013vh,Czakon:2010td,Czakon:2011ve,Czakon:2014oma,Boughezal:2011jf,Cacciari:2015jma,Catani:2007vq,Grazzini:2008tf,Boughezal:2015dva,Gaunt:2015pea,DelDuca:2016csb,DelDuca:2016ily};  refinements of these methods, as well as the development of new proposals, target improved MC stability and efficiency~\cite{Grazzini:2016ctr,Dawson:2016ysj,Caola:2017dug,Caola:2018pxp,Campbell:2017hsw,Moult:2017jsg,Ebert:2018lzn,Herzog:2018ily,Magnea:2018hab}. As a result, exclusive predictions at N$^3$LO are beginning to appear~\cite{Dulat:2017prg,Currie:2018fgr,Cieri:2018oms}.

For numerical stability, it is desirable that the MC generation reflects the structure of the technique for handling IR singularities. The main purpose of this paper is to introduce a method of MC phase space generation that melds particularly naturally with the structure of projection to Born (P2B) subtraction~\cite{Cacciari:2015jma}. The P2B method is used to promote an N$^k$LO inclusive and N$^{k-1}$LO exclusive calculation of a process $X$ and $X+j$, respectively, to a fully exclusive N$^{k}$LO calculation.  The original proposal demonstrated its use to calculate Higgs production in vector boson fusion at NNLO, and it has since been utilised to compute $t$-channel single top production at NNLO~\cite{Berger:2016oht}, and the N$^3$LO corrections to  jet production in deep inelastic scattering~\cite{Currie:2018fgr}. In this paper, we consider the $1\to n$ topology only (example processes being Higgs decay to quarks/gluons, $e^+e^-\to\text{ jets}$, {\it etc.}). The phase space generation we present is not restricted to being used with the P2B method---we highlight this technique as a way of emphasising the very simple way in which the most singular, Born-like IR configurations are generated in the MC program, which can of course be coupled with any subtraction formalism. 

We consider MC generation of phase space in a coordinate system that we summarise  in eqs.~\eqref{eq:ipmv}--\eqref{eq:scaling}, restricting the presentation  to the case  of massless momenta.  The phase space measure is defined in terms of proxy momenta, $q_i$,
\be
d\Phi_n = (2\pi)^{d-n(d-1)} m^{(d-2)n-d}\frac{d\Omega^{d-1}}{2^{d-1}} \frac{d^{d-1}{\bf q}_3}{2|{\bf q}_3|}\cdots \frac{d^{d-1}{\bf q}_n}{2|{\bf q}_n|} \frac{1}{z^{(d-2)n}}   \,,
\label{eq:ipmv}
\ee
with $m^2=P^2$ and where
\be
z^2 = \big(\sum_{i=1}^n q_i\big)^2     \,,
\label{eq:ipmvz}
\ee
and each integration $d^{d-1}{\bf q}_i$ is over all of $\mathbb{R}^{d-1}$.
The  momenta $q_i$ are defined in an energy-scaled Lorentz frame that fixes momenta 1 and 2 such that  $\nobreak{q_1+q_2=(1,{\bf 0})}$,  
\be\begin{aligned}
q_1 = \frac{1}{2}(1,\,{\bf 0},\,-1)  \,, ~~~
q_2 = \frac{1}{2}(1,\,{\bf 0},\,1) \,, ~~~
q_{i\ge 3} = ( |{\bf q}_i| , \,{\bf q}_i)  \,,  
\label{eq:qis}
\end{aligned}
\ee
and the physical momenta, $p_i$, are recovered in a given frame, {\it e.g.}  the centre of mass frame where $P=(m,0,0,0)$, by performing a common scaling transformation, $q_i\to m/z\, q_i$, and the necessary $q_i$ dependent Lorentz transformation, $L$,
\be
p_i = L\left( \frac{m}{z}  q_i\right) \,.
\label{eq:scaling}
\ee

Some nice features of this coordinate system are evident. Most importantly, each momenta $q_i$ is parameterised independently of the others---see eq.~\eqref{eq:qis}---and the limits of integration on the variables $d^{d-1} {\bf q}_3\ldots d^{d-1} {\bf q}_n$ in eq.~\eqref{eq:ipmv} are all  free {\it i.e.} independent of each other. We will see that these facts will simplify the  treatment of IR limits. By construction, the physical momenta, $p_i$, are on-shell and momentum conserving, the latter thanks to the transformation in eq.~\eqref{eq:scaling}. It is trivial to pull out overall Lorentz rotations such that these variables are factorised ({\it e.g.} so as to be treated differently in the MC sampling). 

Eq.~\eqref{eq:ipmv} is, to the best of our knowledge, a new result.  A similar system of independently parameterised momenta variables (IPMVs) does however exist in the literature. It is that employed by the RAMBO MC generator~\cite{Kleiss:1985gy}---the way of treating the phase space measure distinctly from generated physical four-momenta, and the use of the scaling eq.~\eqref{eq:scaling} are, in particular, common features. However, the IPMVs considered here differ significantly in their fixing of momentum 1 and 2; this is crucial for the analysis of the IR limits.

This parameterisation is inspired by a deeper picture of the geometry of phase space in momentum-spinor variables~\cite{hm}.  Specifically, we will make a connection between eq.~\eqref{eq:ipmv} and the well-known Grassmannian geometry of scattering kinematics~\cite{ArkaniHamed:2009dn} (the closely related momentum twistor parameterisation is already utilised in QCD applications, see {\it e.g.}~\cite{Badger:2017jhb,Badger:2018gip}). In doing so, we will anticipate a reparameterisaton of eq.~\eqref{eq:ipmv} to another system of IPMVs in which the most singular, Born-like parton configurations are located around a common origin in parameter space. It is in this sense that MC generation of phase space in these variables reflects the structure of the P2B method, which is concerned with subtracting exactly these singularities.

The organisation of this paper is as follows. We begin by deriving eq.~\eqref{eq:ipmv} in Sec.~\ref{sec:derivation}, in slightly more generality than presented above in that we allow for massive momenta. In Sec.~\ref{sec:grass}, setting $d=4$ and working with massless momenta again, we explore the relation to the Grassmannian, and discover a P2B-friendly change of variables. (Appendix.~\ref{app:grassddim} details a $d$-dimensional version---Grassmannian-esque variables---that are amenable to dimensional regularisation.) In Sec.~\ref{sec:mandies} we study Mandelstam variables, sector decomposition, and the  N$^{n-2}$LO, $1\to n$, IR limits of a sector in terms of these P2B-friendly IPMVs. We use these results in Sec.~\ref{sec:subtract}, exemplifying the use of the IPMVs in conjunction with the  P2B method in a $1\to n$ setting via a toy double real phase space integral. Sec.~\ref{sec:discuss} presents our conclusions.

\section{Derivation of eq.~\eqref{eq:ipmv}, and relation to the Grassmannian}
\label{sec:derivation}

Eq.~\eqref{eq:ipmv} parameterises each point in phase space via proxy momenta, $q_i$, in which $q_1$ and $q_2$ are fixed. The physical momenta are recovered via a locally defined ({\it i.e.} $q_3\ldots q_n$ dependent) Lorentz boost and energy-scaling. The reverse statement is that eq.~\eqref{eq:ipmv} can be obtained from eq.~\eqref{eq:phspcanonical} via a `gauge fixing' (we use this terminology in the sense of coordinate fixing) procedure to set $q_1$ and $q_2$ to the form they take in eq.~\eqref{eq:qis}. Our proof will follow this logic.

We proceed directly from the canonical parameterisation of phase space, eq.~\eqref{eq:phspcanonical}, and use the Fadeev-Popov (FP) trick---see, for example,~\cite{hitoshi}---to perform this gauge fixing. We will work in  general $d$ dimensions, with momenta of arbitrary invariant mass, $m_i$,  specialising to the massless case ({\it i.e.} proving eq.~\eqref{eq:ipmv}) at the end.
In similar notation to that of the RAMBO publication, we write,
\be
p=x L_{\bf b}(q) \,,
\ee
to denote the transformation,
\be\begin{aligned}
p^0 &=x  \gamma (q^0- {\bf b}\cdot {\bf q}) \,,  \\
{\bf p} & = x({\bf q} + \gamma {\bf b}(a \,{\bf b}\cdot {\bf q} - q^0)  )\,,
\end{aligned}
\ee
where $a=\gamma/(1+\gamma)$, and $\gamma =(1-|{\bf b}|^2)^{-1/2}$. 

First, we insert unity into eq.~\eqref{eq:phspcanonical}, 
\be
1&=& \int dx \int d^{d-1} {\bf b}  \,\delta^{(d)}( 1/x L_{\bf -b}(p_{12})  - I ) \left| \frac{ \partial \left(1/xL_{\bf -b}(p_{12}) \right)}{\partial(x,{\bf b })} \right| \,,
\ee
where $p_{12}=p_1+p_2$, and $I=(w,{\bf 0})$, with $w$ an arbitrary scale that represents a choice in the invariant mass of $q_1+q_2$ (and drops out of the final expression). In words, we are integrating over boosts ${\bf b}$ and scalings $x$, with a delta function to implement our gauge fixing condition. 

Eq.~\eqref{eq:phspcanonical} becomes,
\begin{multline}
  (2\pi)^{d-n(d-1)}\int dx \int d^{d-1} {\bf b}\, \delta^{(d)}( 1/x L_{\bf -b}(p_{12})  - I )   \left| \frac{ \partial \left(1/xL_{\bf -b}(p_{12}) \right)}{\partial(x,{\bf b })} \right| \delta^{(d)}(P-\sum_i p_i) \\
  \times \prod_{i=1}^n d^d p_i \,\delta(p_i^2-m_i^2)\theta(p_i^0) \,,
\end{multline}
where we reinstated the delta functions enforcing the on-shell conditions, $p_i^2=m_i^2$, for each particle. We next make the  transformation $p_i \to q_i = 1/x L_{-{\bf b}}(p_i)$. The FP Jacobian must reproduce the Haar measures of the scale and boost transformations, $1/x$ and $\gamma^d$, respectively; the phase space measure is invariant under the boost component of the transformation, but not the scaling, under which it picks up a factor of $x^{dn}$,
\begin{multline}
  (2\pi)^{d-n(d-1)}w^d \int \frac{dx}{x} \int d^{d-1} {\bf b} \, \gamma^{d} \,\delta^{(d)}(q_{1}+q_2  - I)  \, x^{dn} \delta^{(d)}(P -\sum_i  x L_{{\bf b}}(q_i))\\
  \times \prod_{i=1}^n \frac{d^{d-1}{ \bf q}_i}{2 E_i} \,dm_{q_i}^2 \,\delta(x^2m_{q_i}^2-m_i^2) \,.
\end{multline}
We now use the gauge-fixing delta functions to eliminate the integration over the spatial directions of $q_1$, and, going to spherical coordinates, the radial spatial direction of $q_2$, to obtain the usual two particle phase space expression,
\be
\int \frac{d^{d-1}{ \bf q}_1}{2 E_1} \frac{d^{d-1}{\bf q}_2}{2 E_2}  \,\delta^{(d)}(q_{1}+q_2  - I)  = \frac{1}{4} \frac{q_{CM}^{d-3}}{w} d\Omega^{d-1} \,,
\ee
where,
\be
q_{CM}^2=\frac{1}{4w^2} {(w-m_{q_1}-m_{q_2})(w-m_{q_1}+m_{q_2})(w+m_{q_1}-m_{q_2})(w+m_{q_1}+m_{q_2})}\,.
\label{eq:qcm}
\ee
Finally, we can use the momentum conservation delta function to fix the integrals over $dx$ and $d^{d-1}{\bf b}$, after substituting 
\be\begin{aligned}
\delta^{(d)}(P -\sum_{i}  x L_{{\bf b}}(q_i)) &=\frac{1}{x^d} \delta^{(d)}(\frac{1}{x} L_{-{\bf b}}(P) -\sum_{i}  q_i) \,, \\
&= \frac{1}{x^d} m^{-d} \frac{x^{d+1}}{\gamma^d} \delta(x-m/z)\delta^{(d-1)}({\bf b} - {\bf b}_{12}) \,,
\end{aligned}
\ee
where in the second line we have chosen, for simplicity, the centre of mass frame $P=(m,{\bf 0})$, with $z=\sqrt{(\sum_i q_i)^2}$, and  ${\bf b}_{12}= (\sum_i {\bf q}_i)/( \sum_i q^0_i$). We arrive at,
\begin{multline}
  d\Phi_n =(2\pi)^{d-n(d-1)}\frac{1}{4}  m^{(d-2)n-d} w^{d-1} q_{CM}^{d-3} d\Omega^{d-1} \frac{d^{d-1}{\bf q}_3}{2E_3}\cdots \frac{d^{d-1}{\bf q}_n}{2E_n} \frac{1}{z^{(d-2)n}} \\
  \times\left(\frac{m^2}{z^2}\right)^{n}\prod_{i=1}^n dm_{q_i}^2 \,\delta(\frac{m^2m_{q_i}^2}{z^2}-m_i^2) \,,
 \label{eq:generalresult}
\end{multline}
where we can identify the $d\Omega^{d-1}$ as an overall rotation. In the general case, where all particles are massive, the solution of the remaining delta functions is non-trivial; however, significant simplifications occur if we assume the only massive momenta 
are $p_1$ and $p_2$. In this case $z$ does not depend on $m_{q_1}$ and $m_{q_2}$, because it is a function only of $q_1+q_2$: this
is fixed to be $(w,0)$ regardless of masses. Eq.~\eqref{eq:generalresult} then becomes, 
\be
d\Phi_n =(2\pi)^{d-n(d-1)}\frac{1}{4}  m^{(d-2)n-d} w^{d-1} q_{CM}^{d-3} d\Omega^{d-1} \frac{d^{d-1}{\bf q}_3}{2E_3}\cdots \frac{d^{d-1}{\bf q}_n}{2E_n} \frac{1}{z^{(d-2)n}} \,,
\ee
where $q_{CM}$ is given in eq.~\eqref{eq:qcm} with,
\be
m_{q_1}=\frac{z}{m}m_1 \,, \qquad m_{q_2}=\frac{z}{m}m_2 \,. 
\ee
Finally, this reduces to eq.~\eqref{eq:ipmv} for the fully massless case, and where we set $w=1$ (the above expressions are of course independent of $w$, which can be seen by scaling ${ q}_i\to w\, { q}_i$). We will be concerned only with the massless case for the remainder of this paper.

\subsection{Relation to the Grassmannian}
\label{sec:grass}

The boost and scale transformation to fix the special 1-2 frame has its roots in a deeper geometry of phase space: it is associated with a coordinate chart on a Grassmannian manifold.  As well as making the connection with some beautiful aspects of spinor helicity phase space geometry~\cite{hm}, the natural coordinates on the Grassmannian anticipate a coordinate change to the P2B-friendly variables---those which are suited to the generation of Born-like singularities---whose IR limits we will go on to study in Sec.~\ref{sec:mandies}.

To see this, we  proceed in $d=4$ dimensions with massless momenta, naturally represented through spinor-helicity variables,
\begin{equation} \label{eq:spinor-helicity}
  p_i^{\dot{\alpha}\alpha} = \tilde{\lambda}_i^{\dot{\alpha}}\lambda_i^\alpha \,, \qquad p_i^\mu=\frac{1}{2}\,{\sigma}^\mu_{\alpha\dot{\alpha}}p_i^{\dot{\alpha}\alpha}\,,
\end{equation}
where $\tilde{\lambda}^{\dot{\alpha}}=(\lambda^\alpha)^*$ for real momenta. 
Following a similar argument to \cite{ArkaniHamed:2009dn}---see also \cite{hm}---a geometric interpretation of these kinematics is obtained as follows: think of $\lambda_i^\a$ as defining two complex $n$-vectors, $\nobreak{{\bm{\lambda}}^1=(\lambda_1^{\alpha=1},\ldots,\lambda_n^{\alpha=1})}$ and ${\bm{\lambda}}^2=(\lambda_1^{\alpha=2},\ldots,\lambda_n^{\alpha=2})$.  
Momentum conservation (in the centre of mass frame with total momentum $P = \sum_i\tilde{\lambda}_i\lambda_i$) is then the statement that these vectors are orthogonal:
\begin{equation}
{\bm{\lambda}}^1\cdot{\bm{\tilde{\lambda}}}^2=0 \,,\quad {\bm{\lambda}}^1\cdot{\bm{\tilde{\lambda}}}^1=m \,,\quad {\bm{\lambda}}^2\cdot{\bm{\tilde{\lambda}}}^2=m \,.
\end{equation}
$SL(2,\mathbb{C})$ Lorentz transformations act on ${\bm{\lambda}}^1$ and ${\bm{\lambda}}^2$ as transformations within the $({\bm{\lambda}}^1,{\bm{\lambda}}^2)$ plane. 
Hence, massless $n$-particle phase space, modulo overall spatial rotations, can be described as the set of all planes passing through the origin in $\mathbb{C}^n$. 
This space is the complex Grassmannian $Gr(2,n)$.

In order to utilise this geometric interpretation of phase space, we first need to define coordinates on the Grassmannian. 
In describing each point in $Gr(2,n)$ by two arbitrary, linearly-independent vectors there is a redundancy associated with $GL(2,\mathbb{C})$ transformations acting in the plane. 
In the case of interest this is nothing other than the $SL(2,\mathbb{C})$ of the Lorentz transformations, and a complex scale transformation. 
This redundancy can be removed via a gauge fixing, with each gauge choice corresponding to a different chart on the Grassmannian. 
A particularly convenient choice that we will adopt is,
\begin{equation} \label{eq:gauge_fix}
  \begin{aligned}
    {\bm{\lambda}}^1 &= e^{-i\phi}(0,1,u_3,\ldots,u_n)\,, \\
    {\bm{\lambda}}^2 &= ~~e^{i\phi}(1,0,v_3,\ldots,v_n) \,.
  \end{aligned}
\end{equation}
In fact, this is a partial gauge fixing up to the additional phase $\phi$ which acts on the spinors as a Lorentz rotation.

At this point we recall that there is a further redundancy that arises when using spinor variables, associated with the fact that the momenta $p_i^\mu$ are invariant under the transformation $\lambda_i^\alpha \to e^{i\theta_i}\lambda_i^\alpha$. 
This redundancy can be used to remove a further $n-1$ phases (one common phase is already removed by the $GL(2,\mathbb{C})$ gauge-fixing), such that 
\begin{equation}
  \begin{aligned}
    {\bm{\lambda}}^1 &= (0,1,u_3e^{i(\phi_3-\phi)},\ldots,u_ne^{i(\phi_n-\phi)})\,, \\
    {\bm{\lambda}}^2 &= (1,0,v_3,\ldots,v_n) \,,
  \end{aligned}
  \label{eq:phases}
\end{equation}
where we have moved to polar coordinates with $0\le u_i, v_i<\infty$ and $0\le \phi_i <2\pi$. 
Note that we could have  used  the remaining gauge freedom in $\phi$ to remove one of the $\phi_i$; however, since phase space is the product of the Grassmannian and the Lorentz rotations it is convenient for now to retain this additional angle and simply redefine $\phi_i\to\phi_i+\phi$. 
This leaves us with $3n-6$ real parameters which, combined with the remaining two Lorentz rotations, is precisely the dimension of $n$-particle phase space. 

The Grassmannian picture also allows us to directly obtain the phase space integration measure.
Using the fact that the Grassmannian is a K\"ahler manifold via its Pl\"ucker embedding in $\mathrm{CP}^{{{n}\choose{2}} -1}$, we can derive (see appendix~\ref{app:measure} for details) an expression for the phase space measure in these coordinates:
\begin{equation}
  d\Phi_n = (2\pi)^{4-3n} \frac{d\Omega^3}{2^3}\frac{1}{{z^{2n}}}\prod_{i=3}^n u_i v_i\, du_i dv_i d\phi_i \,,
  \label{eq:uvipmv}
\end{equation}
where $z$ is given in terms of $u$ and $v$ in eq.~\eqref{eq:appz} (or equivalently using eq.~\eqref{eq:ipmvz} with the $q_i$ given in eq.~\eqref{eq:qis2} below). 
A dimensionally continued version of eq.~\eqref{eq:uvipmv}---Grassmannian-esque phase space---is presented in appendix~\ref{app:grassddim}, along with a discussion of its associated geometry.

\section{P2B-friendly IPMVs}
\label{sec:mandies}

In exploring the geometric origins of fixing two of the phase space momenta ($q_1$ and $q_2$) to canonical values, we ended up with a measure eq.~\eqref{eq:uvipmv} in terms of Grassmannian coordinates. That this is a simple change of variables from eq.~\eqref{eq:ipmv} can be most easily seen by reconstructing the four-momenta from the spinors in eq.~\eqref{eq:phases}, via eq.~\eqref{eq:spinor-helicity}. We have,
\be
\begin{aligned} \label{eq:qis2}
  q^\mu_1 &= \frac{1}{2} (1,\,0,\,0,\,-1) \,, ~~~q^\mu_2 = \frac{1}{2} (1,\,0,\,0,\,1) \,,  \\
  q^\mu_{i\geq3} &= \frac{1}{2} (u_{i}^2+v_{i}^2,\, 2 u_{i} v_{i}\cos\ph_{i},\, 2 u_{i} v_{i} \sin\ph_{i},\, u_{i}^2-v_{i}^2) \,,
\end{aligned}
\ee
where $0\le u_i, v_i<\infty$ and $0\le \phi_i <2\pi$, and where, as discussed under eq.~\eqref{eq:phases}, one combination of the $\ph_i$ can be identified as a further overall rotation (in practice one can set {\it e.g.} $\phi_3=0$). 
As in eq.~\eqref{eq:qis}, these are not the physical momenta, but must be related to the $p_i$ via eq.~\eqref{eq:scaling}; we will work with $m=1$ in the following.
In this section we will see why this change of variables from eq.~\eqref{eq:qis} to eq.~\eqref{eq:qis2} is P2B-friendly.

We are going to be discussing IR limits, and any collinear or soft configuration will stay collinear or soft under any {\it finite} scaling, boost or rotation. However, it is clear that to reconstruct  a configuration of physical momenta where $p_1$ and $p_2$ are collinear to each other and/or soft, an {\it infinite} boost and/or scaling are going to be required, associated with  limits where one or more of the $u_i$ or $v_i$ tend to infinity; these are not straightforward  to analyse. This motivates the splitting of the total phase space into sectors, such that  eq.~\eqref{eq:qis2}  parameterises the sector where $s_{12}$ remains finite, thus precluding the tricky projective IR limits. With such a sector splitting, we see one of the reasons IPMVs are well suited to IR subtraction: the complicated, non-linear transformations from $q_i$ to $p_i$ (which, in addition, undo the independent nature of the parameterisation) do not enter the discussion of IR singularities.

\subsection{Mandelstam variables and sector decomposition}

From eq.~\eqref{eq:qis2} (and the scaling component of eq.~\eqref{eq:scaling} to the physical momenta), it is straightforward to see that the Mandelstam invariants in these variables take three different forms,
\be
&&s_{12} = \frac{1}{z^2}  \,,  \label{eq:mand0}\\
&&s_{1i} = \frac{u_i^2}{z^2}  \,, ~~~ s_{2i} = \frac{v_i^2}{z^2}  \,, ~~~~~\,i \ge 3 \,,  \label{eq:mand1}\\
&&s_{ij} = \frac{u_i^2 v_j^2 + u_j^2  v_i^2 - 2 u_i v_i u_j v_j \cos(\ph_i- \ph_j) }{z^2}  \,,  \ ~~~~~ 3\le i < j \le n  \,.  \label{eq:mand2}
\ee
Sectors are defined through requiring one Mandelstam invariant to be greater than all others. As already mentioned, the parameterisation above, with momenta 1 and 2 singled out as special, is most naturally associated with the 1-2 sector,
\be
s_{12} > \text{all other } s_{ij} \,.
\ee
The strategy for obtaining simple Born-like IR limits everywhere in phase space will be to parameterise each different sector $i$-$j$ via a permutation of the above IPMVs, such that momenta $i$ and $j$ are the special momenta. From the discussion of the previous section, this is precisely choosing a different coordinate chart on the Grassmannian for each different sector. For the remainder of this section, we continue to work in the 1-2 parameterisation without loss of generality.

We need to define sectors in terms of the IPMVs.  For this we can inspect the form of the Mandelstam invariants.  All invariants have a common factor of $1/z^2$; comparing eq.~\eqref{eq:mand0} to eq.~\eqref{eq:mand1}, we get the constraint,
\be
0\le u_i < 1 \,,~~~~0\le v_i < 1  \,, ~~~~~ \,i\ge 3 \,.
\label{eq:limit1}
\ee
For $n>3$, part of this hypercube must be cut out; comparing eq.~\eqref{eq:mand0} to eq.~\eqref{eq:mand2} we have,
\be
u_i^2 v_j^2 + u_j^2 v_i^2 - 2 u_i v_i u_j v_j \cos(\ph_i- \ph_j) <1  \,,~~~ 3\le i< j \le n \,,
\label{eq:limit2}
\ee
which is easily implemented numerically in an MC program.

\subsection{The N$^{n-2}$LO IR limits of a sector}
\label{sec:irlimits}

N$^{n-2}$LO IR limits occur in the regions of phase space approaching two-particle kinematics. We now study these limits with P2B-friendly IPMVs. 

Within the 1-2 sector, two-particle kinematics can only be achieved with a configuration where momenta 1 and 2 are back-to-back, and each particle $3$ through $n$ is collinear with either particle 1 or 2, or else soft.  For particle $i$ to be collinear to particle 1 (2), we set $u_i\to0$ ($v_i\to0)$; for particle $i$ to be soft, both $u_i,v_i\to0$. These limits do not involve the angular variables $\ph_i$. 

That is, all N$^{n-2}$LO of sector 1-2 are intersections of subsets of zero hyper-surfaces of the hypercube $0\le u_i,v_i< 1$. From the point of view of simplicity and MC efficiency, this is an attractive feature. An example: the intersections of zero hyper-surfaces and the corresponding N$^{n-2}$LO limits in the 1-2 sector for the case $n=4$ are given by,
\be
\begin{array}{r|ccc}
        \text{Zero hyper-surfaces}  ~     &~~~~u_3\to0~~~~~~&~~~~~v_3\to 0~~~~~ &~~~~~ u_3,v_3\to0~~~~~\\\hline
u_4\to 0~ & 1||3||4 &  2||3\text{ and }1||4 & 1||4\text{ and }3\text{ soft} \\
v_4\to 0 ~&   1||3\text{ and }2||4   & 2||3||4&  2||4\text{ and }3\text{ soft} \\  
u_4,v_4\to 0 ~&~~ 1||3\text{ and }4\text{ soft}~~   &~~2||3\text{ and }4\text{ soft} ~~& 3,4\text{ soft}
\end{array}
\ee

As detailed further in the next section, the P2B method,  applied to $1\to n$ processes, assumes that the 3-jet rate is known at N$^{n-3}$LO; it then provides the prescription to subtract off the remaining N$^{n-2}$LO IR singularities.
Here we see that the limit  $u_i,v_i\to0$ for all $i$ provides a simple projection that is used in the P2B method: all the possible N$^{n-2}$LO IR configurations in this sector have the same 2-jet kinematics. This is the reason why we claim the IPMVs of eq.~\eqref{eq:qis2} are P2B-friendly.

\section{P2B subtraction of $1\to n$ processes with IPMVs}
\label{sec:subtract}

The P2B method for observables of $1\to n$ processes in massless QCD is implemented schematically as (we adopt the notation of~\cite{Currie:2018fgr}),
\be
\frac{d\sigma^{N^{n-2} LO}_X}{d\mathcal{O}}  = \frac{d\sigma^{N^{n-3} LO}_{X+j}}{d\mathcal{O}} - \frac{d\sigma^{N^{n-3} LO}_{X+j}}{d\mathcal{O}_B}+ \frac{d\sigma^{N^{n-2} LO,\,\text{incl.}}_{X}}{d\mathcal{O}_B} \,,
\label{eq:p2b}
\ee
where $d\mathcal{O}$ defines the differential IR-safe observable with full kinematics, and  $d\mathcal{O}_B$ defines the Born kinematics onto which the full event is projected. In this case $d\mathcal{O}_B=d\Phi_2$. Based on the discussion of the previous section, we can now give our prescription using the IPMVs. 

First we partition the  phase space of the full event into sectors. Within each sector $i$-$j$, we parameterise using eq.~\eqref{eq:uvipmv}, with momenta $q_i$ and $q_j$ taking the special fixed values, {\it i.e.} a permutation of eq.~\eqref{eq:qis2}. Let us focus on the sector 1-2. The full event kinematics are given by the $q_i$ of eq.~\eqref{eq:qis2}, scaled and boosted to the physical $p_i$. The projected kinematics can be taken as the two-jet configuration that coincides with all N$^{n-2}$LO IR limits, as per the discussion of Sec.~\ref{sec:irlimits}: $p_1^B=\frac{1}{2}(1,0,0,-1)$, $p_2^B=\frac{1}{2}(1,0,0,1)$.

\subsection{NNLO double real toy example}
We will illustrate the P2B method using IPMVs with a double real toy phase space integral, chosen so as to provide a straightforward differential cross-check (see below),
\be
\frac{d\sigma^{NNLO}_{RR}}{d\mO} = \int d\Phi_4 \,\frac{s_{12}}{s_{13}s_{23}s_{14}s_{24}}  \, J(\mO_4)\,.
\label{eq:nnloex}
\ee

We construct the NLO 3 jet calculation using the Catani-Seymour (CS) dipole formalism~\cite{Catani:1996vz}. That is,
\be
\frac{d\sigma^{NLO}_{RR+ j}}{d\mO} = \int d\Phi_4 \bigg(  d\s_{RR} J(\mO_4) - d\s_{C} J(\mO_{3,\,\text{dip}})  \bigg) + \int d\Phi_3  \, d\s_{IC} J(\mO_3)  \,,
\ee
where $d\s_{RR}$ is the  function of Mandelstam variables in eq.~\eqref{eq:nnloex}, and $d\s_{C}$ is the  CS dipole counter-term; the jet function $J(\mO_{3,\,\text{dip}})$ is evaluated with the dipole momenta  constructed using the usual dipole map. The integrated counter-term is denoted $d\s_{IC}$.

For the inclusive NNLO calculation we perform the reduction of eq.~\eqref{eq:nnloex} to the master integrals in Ref.~\cite{Gehrmann-DeRidder:2003pne}  using the
reverse unitarity procedure of Ref.~\cite{Anastasiou:2002yz}, and the program FIRE~\cite{Smirnov:2008iw,Smirnov:2013dia}. We find, in $d=4-2\e$ dimensions,
\be
 \frac{d\sigma^{NNLO,\,\text{incl}}_{RR}}{d\mO_B}  =  \frac{d\sigma^{NNLO,\,\text{incl}}_{RR}}{d\Phi_2} =\frac{(4\pi)^{-4+2\e}}{\Gamma(1-\e)^2}   \bigg( \frac{1}{\e^4}-\frac{4\pi^2}{3\e^2}-\frac{24\zeta_3}{\e} +\frac{8\pi^4}{45}+O(\e) \bigg) \,.
\ee

These ingredients can now be used to obtain the exclusive NNLO calculation eq.~\eqref{eq:nnloex}, via the P2B prescription of eq.~\eqref{eq:p2b},
\be\begin{aligned}
\frac{d\sigma^{NNLO}_{RR}}{d\mO} = ~~& \int d\Phi_4 \bigg(  d\s_{RR} J(\mO_4) - d\s_{C} J(\mO_{3,\,\text{dip}})  - d\s_{RR} J(\mO_{4\to B}) + d\s_{C} J(\mO_{3,\,\text{dip}\to B}) \bigg)  \\
 +& \int d\Phi_3 \bigg(   d\s_{IC} J(\mO_3)  -   d\s_{IC} J(\mO_{3\to B})\bigg)  \\
 +& \frac{d\sigma^{NNLO,\,\text{incl}}_{RR}}{d\mO_B}  \,.
\end{aligned}
\label{eq:p2bex}
\ee
The IPMVs introduced in this work are used to construct  the phase space and the relevant projections to Born kinematics. Explicitly, the four particle phase space is split into sectors. For sector 1-2, we parameterise phase space using eq.~\eqref{eq:uvipmv} (absorbing $\phi_3$ as an overall rotation, $d\Omega^2$, and writing $\phi_4=\phi$),
\be\begin{aligned}
d\Phi_4  &=  (2\pi)^{-6}  d\Phi_2 \,d\Omega^2\,\frac{u_3 v_3 u_4 v_4} {z^8}  \, du_3 dv_3 du_4 dv_4 d\ph\,,
\end{aligned}
\label{eq:exphsp}
\ee
where,
\be
z^2=1+u_3^2+v_3^2 + u_4^2 + v_4^2 + u_3^2 v_4^2 + u_4^2 v_3^2 - 2 u_3 v_3 u_4 v_4 \cos\ph \,,
\ee
and the limits of integration are, from eqs.~\eqref{eq:limit1},~\eqref{eq:limit2},
\be
0\le u_i,v_i\le 1\,,~~~~0\le\ph<2\pi \,,~~~~\text{with}~~u_3^2 v_4^2+ u_4^2v_3^2 - 2 u_3 v_3 u_4 v_4 \cos \ph<1 \,.
\ee
The full-event kinematics, $p_1,\ldots,p_4$, are constructed first by obtaining the proxy momenta $q_i$ via eq.~\eqref{eq:qis2},
\be\begin{aligned}
&q_1=\frac{1}{2}(1,0,0,-1)\,,~~q_2=\frac{1}{2}(1,0,0,1)\,,  \\ &q_3=\frac{1}{2}(u_3^2+v_3^2,\,2u_3v_3,\,0,\,u_3^2-v_3^2)\,,\\ ~~&q_4=\frac{1}{2}(u_4^2+v_4^2,\,2u_4v_4\cos\ph,\,2u_4v_4\sin\ph,\,u_4^2-v_4^2)\,,
\end{aligned}\label{eq:qisex}\ee
and then using eq.~\eqref{eq:scaling} to recover the $p_i$,
\be
p_i = L\left( \frac{1}{z}  \,q_i\right) \, ,
\ee
where $L$ is the ($u$, $v$, $\ph$)-dependent Lorentz boost to the CM frame of the $p_i$.  The projection to Born is taken to be the limit $u_i,v_i\to0$ of the above kinematics, giving, 
\be
p_1^B=\frac{1}{2}(1,0,0,-1),~~~ p_2^B=\frac{1}{2}(1,0,0,1) \,;
\label{eq:p2bmomenta}
\ee
these are the momenta that are used in the jet function $J(\mO_{4\to B})$. Because we have the full-event kinematics $p_i$ as explicit functions of the IPMVs, we simply follow the standard CS prescription for constructing the dipole momenta out of the $p_i$;  the limit $u_i,v_i\to0$ also projects the dipole momenta to those of  eq.~\eqref{eq:p2bmomenta}. Different sectors are treated via an index permutation of eq.~\eqref{eq:qisex}. The   
 analogous construction is used for the three particle MC integral in eq.~\eqref{eq:p2bex}.

We constructed an independent check of the differential distributions obtained using the P2B method with IPMVs. The integrand of eq.~\eqref{eq:nnloex} in the  Grassmannian-esque variables given in appendix~\ref{app:grassddim} has factorised limits as $u_i,v_i\to 0$; furthermore these limits comprise all IR singularities, owing to the $s_{12}$ in the numerator. We can, then, employ a straightforward plus-prescription limit subtraction to obtain the distributions.

\section{Discussion}
\label{sec:discuss}

In this paper we introduced a system of independently parameterised momenta variables---defined through eqs.~\eqref{eq:uvipmv},~\eqref{eq:qis2}, and~\eqref{eq:scaling}---that form the basis for an MC generation of phase space. We explored its geometric (Grassmannian) origins. The construction is general in that it applies to $1\to n$ body phase space for any $n$, and it is designed to efficiently and simultaneously parameterise all N$^{n-2}$LO IR singularities that occur in a given sector. As shown in Sec.~\ref{sec:irlimits}, these singularities are intersections of various zero hypersurfaces of a hypercube in parameter space. We exemplified the use of these IPMVs in conjunction with the P2B technique by considering a toy NNLO example in Sec.~\ref{sec:subtract}. 
Of course, it is important in future to test the efficiency of using the IPMVs in conjunction with the P2B method (or other scheme) in interesting phenomenological $1\to n$ settings at NNLO and beyond.

The generality of the IPMV parameterisation derived in Sec.~\ref{sec:derivation} suggests multiple avenues for further exploration. For instance, it is natural to investigate its use in combination with phase space recursion and amplitude factorisation, with an eye to a generic subtraction scheme at NNLO. Another logical direction is to consider topologies beyond $1\to n$ and/or its use in higher order calculations that involve massive particles. 

Finally, we focused here on numerical MC techniques, but it will be interesting to explore the analytic use of these variables; Appendix~\ref{app:grassddim} presents some initial steps in this direction.

\section*{Acknowledgements}
We would like to thank  Keith Hamilton, Brian Henning, Franz Herzog, Hitoshi Murayama, and Giulia Zanderighi for useful discussions, and Simon Badger and Brian Henning for comments on a draft version of this manuscript. This work is supported by the World Premier International Research Center Initiative (WPI), MEXT, Japan. TM is supported by JSPS KAKENHI Grant Number JP18K13533. TM is grateful to the Mainz Institute for Theoretical Physics (MITP) for its hospitality as this work was being finalised.

\appendix

\section{Grassmannian Phase Space Measure}
\label{app:measure}

The Grassmannian $Gr(2,n)$ can be embedded into the complex projective space $\mathrm{CP}^{{{n}\choose{2}} -1}$ and hence is a K\"ahler manifold. The corresponding K\"ahler potential is
\begin{equation} \label{eq:Kahler_potential}
  f_\mathbb{G} = \log\det(ZZ^\dagger) \,,
\end{equation}
where $Z$ is the $2 \times n$ matrix of coordinates on the Grassmannian, which is defined up to a $GL(2,\mathbb{C})$ transformation. 
Following our gauge choice in Eq.~\eqref{eq:gauge_fix} we have
\begin{equation}
  Z = 
  \begin{pmatrix}
    0 & 1 & u_3 & \cdots & u_n \\
    1 & 0 & v_3 & \cdots & v_{n}
  \end{pmatrix} \,.
\end{equation}
Evaluating the determinant in the K\"ahler potential yields
\begin{equation}
  \det(ZZ^\dagger)= \left( 1+\sum_{i=3}^n\bar{u}_iu_i \right) \left( 1+\sum_{i=3}^n\bar{v}_iv_i \right) - \left( \sum_{i=3}^n\bar{u}_iv_i \right) \left( \sum_{i=3}^n\bar{v}_iu_i \right) \equiv z^2\,.
\label{eq:appz}
\end{equation}
This is precisely the same $z^2$ that appears in Eq.~\eqref{eq:ipmvz}, up to additional (redundant) phases. 
From the above K\"ahler potential, and its corresponding K\"ahler form $\omega_\mathbb{G}=i/2\,\partial\bar{\partial}f_\mathbb{G}$, it is reasonably straightforward to obtain the measure on the Grassmannian,
\begin{equation}
  \mathrm{vol}_\mathbb{G} = \omega_\mathbb{G}^{2(n-2)} = \left(\frac{i}{2}\right)^{2(n-2)} \frac{1}{z^{2n}}\,\prod_{i=3}^n du_i d\bar{u}_i dv_i d\bar{v}_i \,,
\end{equation}
where $\omega_\mathbb{G}^{2(n-2)}$ is understood as a wedge product. Transforming from holomorphic to polar coordinates ($u_i \to u_ie^{i\phi_i}$, $v_i \to v_ie^{i\theta_i}$), removing the redundant phases (which we take to be $\theta_i$), and including the integration over the Lorentz rotations and overall normalisation, we obtain the expression for the differential phase space volume in Grassmannian coordinates,
\begin{equation}
  d\Phi_n = (2\pi)^{4-3n} \frac{d\Omega^3}{2^3}\frac{1}{{z^{2n}}}\prod_{i=3}^n u_i v_i\, du_i dv_i d\phi_i \,.
\end{equation}

\section{Dimensional regularisation with Grassmannian-esque variables}
\label{app:grassddim}

The Grassmannian exposition suggests writing,
\be
d^{d-1}{\bf q}_i &=& dq^z_i\,d^{d-2}{\bf q}^{||}_i \,,\\ 
&=& dq^z_i |{\bf q}^{||}_i|^{d-3}\,d|{\bf q}^{||}_i|  \, d\Omega^{d-2}_i \,,
\ee
and making the change of variables,
\be
q^z_i = \frac{1}{2}(u_i^2-v_i^2) ,~~~~~|{\bf q}^{||}_i| = u_i \,v_i \,, ~~~~~0\le u_i, v_i < \infty,
\label{eq:uvchange}
\ee
such that eq.~\eqref{eq:ipmv} can be cast in the form
\be
d\Phi_n = (2\pi)^{d-n(d-1)}m^{(d-2)n-d}\frac{d\Omega^{d-1}}{2^{d-1}} \frac{1}{z^{(d-2)n}}   \prod_{i=3}^n (u_iv_i)^{d-3} d u_i dv_id\Omega^{d-2}_i  \,.
\ee
We work out some examples in $d=4-2\e$. 
For two-particle phase space, one simply obtains,
\begin{equation}
  d\Phi_2 = 2^{-3+2\e}(2\pi)^{-2+2\e} m^{-2\e} d\Omega^{3-2\e} \,.
\end{equation}
For three-particle phase space, the $d\Omega_3^{d-2}$ can be identified as the additional overall rotation,
\begin{equation}
  d\Phi_3 = 2^{-3+2\e}(2\pi)^{-5+4\e} m^{2-4\e} d\Omega^{3-2\e} d\Omega^{2-2\e}  \left(\frac{u_3^2v_3^2}{z^6}\right)^{-\epsilon}\frac{u_3v_3}{z^6} \, du_3dv_3 \,.
\end{equation}
For four-particle phase space, again we can identify $d\Omega_3^{d-2}$ as an overall rotation; we then write $d\Omega_4^{d-2}=\sin^{d-4}\th\,d\th d\Omega^{d-3}$ where $0\le \th<\pi$,
\begin{equation}
  d\Phi_4 =  2^{-3+2\e} (2\pi)^{-8+6\e} m^{4-6\e} d\Omega^{3-2\e} d\Omega^{2-2\e}  d\Omega^{1-2\e}\left(\frac{u_3^2v_3^2u_4^2v_4^2\sin^2\th}{z^8}\right)^{-\epsilon}\frac{u_3 v_3 u_4 v_4 }{z^8} \, du_3dv_3du_4dv_4d\th \,.
\end{equation}
In the $\e=0$ limit, $d\Omega^{1}d\th$ can be written as $d\ph$ with $0\le \ph<2\pi$; this identifies with the variable $\phi$ in eq.~\eqref{eq:exphsp}, that appears in the four-momenta as per eq.~\eqref{eq:qisex}.

We emphasise that the $u$, $v$ variables introduced in eq.~\eqref{eq:uvchange} do not have anything {\it a priori} to do with spinor variables, in contrast to the $u$ and $v$ in eq.~\eqref{eq:uvipmv} which have a direct relation with the Grassmannian geometry. They are, however, suggestive of a  way of associating a dimensional continuation to the geometry described in terms of the complex $n$-vectors ${\bm{\lambda}}^1$ and ${\bm{\lambda}}^2$ defined under eq.~\eqref{eq:spinor-helicity}. This is most easily appreciated through the dependence of the phase space volume on the dim reg parameter $\e$, for $d=4-2\e$, as written in the form given in~\cite{Gehrmann-DeRidder:2003pne},
\be
\text{Vol}(\Phi_n^{4-2\e}) \propto \frac{\Gamma(1-\e)^n}{\Gamma(n(1-\e))\Gamma( (n-1)(1-\e))}  \,.
\label{eq:phspevol}
\ee
We write,
\be
du_i u_i^{d-3}d\Omega^{d-2}_i &=& d^{d-2} {\bf u}_i  \,,\\
dv_i v_i^{d-3}  &=& dv_i v_i^{d-3} \frac{d\Omega^{d-2}}{S^{d-2}} = d^{d-2} {\bf v}_i / S^{d-2}\,,
\label{eq:pseudolittle}
\ee
where 
\be
S^{d} = \int d\Omega^{d} = \frac{2\pi^{\frac{d}{2}}}{\Gamma(\frac{d}{2})} \,.
\ee
Using this, we can derive that, 
\be
\prod_{i=1}^n \frac{d^{d-1} {\bf q}_i}{2E_i} = \prod_{i=1}^n (u_i v_i)^{d-3} du_i dv_i d\Omega^{d-2}_i = \frac{1}{(S^{d-2})^n} d^{n(d-2)} {{{\bm \lambda}}^1} \,d^{n(d-2)} {{{\bm \lambda}}^2} \,.
\label{eq:dimcont}
\ee
In $d=4$, the $2n$-real dimensional integral $d^{2n} {{{\bm \lambda}}}$ corresponds to integrating over the complex $n$ dimensional vector ${{{\bm \lambda}}}$.
We see that, in moving from $d=4$ to $d=4-2\e$, the $2n$ real-dimensional vectors ${{{\bm \lambda}}^1}$ and ${{{\bm \lambda}}^2}$ become $2n(1-\e)$ dimensional, {\it i.e.} the dimensions  of these vectors get squashed by a factor of $(1-\e)$. The volume of the Grassmannian contains the product of the volume of a complex $n$ and $(n-1)$ sphere in particle number space (see~\cite{hm}). Here, we see that the $\Gamma(n(1-\e))$ and $\Gamma( (n-1)(1-\e))$ factors in eq.~\eqref{eq:phspevol} arise geometrically from squashing these balls by this factor of $(1-\e)$.  The geometric origin of the  $\Gamma(1-\e)^n$ is the analytic continuation of the volume of the space that the little group, $SO(d-2)$, acts on, as introduced in eq.~\eqref{eq:pseudolittle} for each particle. We will further explore these ideas elsewhere.

\bibliographystyle{jhep}
\bibliography{bibliography}

\end{document}